\documentclass[graybox]{svmult}

\usepackage{mathptmx}       
\usepackage{helvet}         
\usepackage{courier}        
\usepackage{type1cm}        
\usepackage{makeidx}         
\usepackage{graphicx}        
\usepackage{multicol}        
\usepackage[bottom]{footmisc}

\makeindex             

\begin{document}

\title*{The charming beauty of the strong interaction}
\author{Laura Tolos}
\institute{Laura Tolos \at Institut f\"ur Theoretische Physik, University of Frankfurt, Max-von-Laue-Str. 1, 60438 Frankfurt am Main, Germany,
          \at Frankfurt Institute for Advanced Studies,  University of Frankfurt, Ruth-Moufang-Str. 1, 60438 Frankfurt am Main, Germany,
          \at Institute of Space Sciences (CSIC-IEEC), Campus Universitat Aut\`onoma de Barcelona, Carrer de Can Magrans, s/n, 08193 Cerdanyola del Vall\`es,
Spain, \\
\email{tolos@th.physik.uni-frankfurt.de}}

%
%
\maketitle

\abstract*{Charmed and beauty hadrons in matter are discussed within a unitarized coupled-channel model consistent with heavy-quark spin symmetry.  We analyse the formation of $D$-mesic states  as well as the propagation of charmed and beauty hadrons in heavy-ion collisions from LHC to FAIR energies.}

\abstract{Charmed and beauty hadrons in matter are discussed within a unitarized coupled-channel model consistent with heavy-quark spin symmetry.  We analyse the formation of $D$-mesic states  as well as the propagation of charmed and beauty hadrons in heavy-ion collisions from LHC to FAIR energies.}

\section{Introduction}
One of the main research activities in nuclear and particle physics is the exploration of the Quantum Chromodynamics (QCD) phase diagram for high density and/or temperature. Up to now the studies have been concentrated for light quarks due to energy constraints of the experimental setups, but with the upcoming research facilities, the goal is to move to the heavy-quark domain, where heavy degrees of freedom, such as charm and beauty, play a crucial role.

In order to understand the QCD phase diagram, one needs first to understand the interaction between heavy hadrons. In particular, the nature of newly discovered heavy excited states is of major concern, whether they can be described within the standard quark model and/or better understood as dynamically generated states via hadron-hadron interactions.

Given the success of unitarized coupled-channel approaches in the description of some of the existing experimental data in the light-quark sector,  charmed and beauty degrees of freedom have been recently incorporated in these models and several experimental states have been described as  excited baryon molecules. Some examples can be found in Refs.~\cite{Tolos:2004yg, Tolos:2005ft, Lutz:2003jw, Lutz:2005ip, Hofmann:2005sw, Hofmann:2006qx, Lutz:2005vx, Mizutani:2006vq, Tolos:2007vh, JimenezTejero:2009vq, Haidenbauer:2007jq, Haidenbauer:2008ff, Haidenbauer:2010ch, Wu:2010jy, Wu:2010vk, Wu:2012md, Oset:2012ap,Liang:2014kra,Yamaguchi:2013ty,Lu:2014ina,Liang:2014eba,Roca:2015dva,Lu:2016gev,Hosaka:2016ypm,Montana:2017kjw,Debastiani:2017ewu}. However, some of these models are not fully consistent with heavy-quark spin symmetry (HQSS) ~\cite{Isgur:1989vq}, which is a  QCD symmetry that appears when the quark masses become larger than the typical confinement scale. Thus, a model that incorporates HQSS constraints has been developed in the past years \cite{GarciaRecio:2008dp,Gamermann:2010zz,Romanets:2012hm,GarciaRecio:2012db,Garcia-Recio:2013gaa,Tolos:2013gta,Garcia-Recio:2015jsa,Nieves:2017jjx}.

Once the interaction between heavy hadrons has been determined, the study of the properties of heavy hadrons in nuclear matter requires the inclusion of nuclear medium modifications \cite{Rapp:2011zz}. In this way, it is possible to study the formation of heavy mesic states in nuclei \cite{GarciaRecio:2010vt,GarciaRecio:2011xt,Yamagata-Sekihara:2015ebw}, as well as the propagation of charmed and beauty in heavy-ion collisions \cite{Tolos:2013kva,Torres-Rincon:2014ffa,Ozvenchuk:2014rpa,Song:2015sfa,Song:2016rzw}, all of these topics matter of the present paper.

\section{Charm under Extreme Conditions}
\label{charm}
\subsection{Excited charmed baryons}
\label{excitedcharm}

Recently a predictive model has been developed for four flavors including all ground-state hadrons (pseudoscalar and vector mesons, and $1/2^+$ and $3/2^+$ baryons). This scheme reduces to the Weinberg-Tomozawa (WT) interaction when Goldstone bosons are involved and includes HQSS in the sector where heavy quarks appear. In fact, this model is justified due to the results of the SU(6) extension in the three-flavor sector \cite{Gamermann:2011mq} and is based on a formal plausibleness on how the interactions between heavy pseudoscalar mesons and baryons emerge in the vector-meson exchange picture. The WT potential can be then used to solve the on-shell Bethe-Salpeter equation in coupled channels so as to calculate the scattering amplitudes. The poles of the scattering amplitudes are the dynamically-generated charmed baryonic resonances (see Ref.~\cite{Tolos:2013gta} for a review). 

Dynamically generated states with different charm and strangeness  are predicted in Refs.~\cite{GarciaRecio:2008dp,Gamermann:2010zz,Romanets:2012hm}. The studies are constrained to the states coming from the most attractive representations of the SU(6) $\times$ HQSS scheme. Some of them can be
identified with known states from the PDG \cite{Patrignani:2016xqp}, by comparing the PDG data on these states with the mass, width and the dominant couplings to the meson-baryon channels. 

In this work, as an example, the results in the $C=1,S=0, I=0$ are presented.  In Ref.~\cite{Romanets:2012hm} three $\Lambda_c$ and one $\Lambda_c^*$ are obtained. A pole around 2618.8 MeV is identified with the experimental $\Lambda_c(2595)$ resonance, while a second broad $\Lambda_c$
state at 2617~MeV shows a similar two-pole pattern as in the $\Lambda(1405)$ case \cite{Jido:2003cb}, coupling strongly to $\Sigma_c \pi$. The third spin-$1/2$ $\Lambda_c$ around 2828~MeV cannot be assigned to any experimental state.  With regards to the spin-$3/2$ $\Lambda_c$ state, this is assigned to  $\Lambda_c(2625)$. 

\begin{figure} [t]
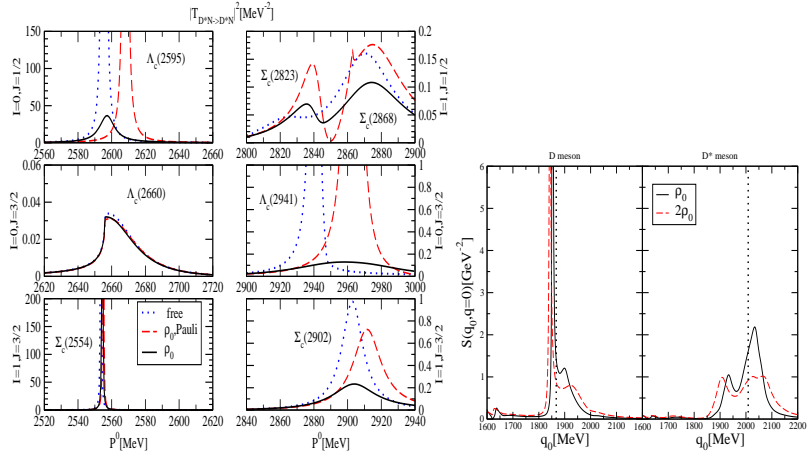

\begin{center} 
\includegraphics[width=0.5\textwidth,height=6cm]{art_reso2.eps}
\includegraphics[width=0.4\textwidth,height=4cm]{art_spec.eps}
\caption{ Left: Charmed baryonic resonances in dense matter (taken from Ref.~\cite{Tolos:2009nn}). Right: The $D$ and $D^*$ spectral functions in dense nuclear matter at zero momentum (taken from Ref.~\cite{Tolos:2009nn}). }
\label{fig2}
\end{center}
\end{figure}

 \subsection{Charmed hadrons in matter}

The inclusion of dense matter effects modifies the mass and width of the dynamically-generated baryonic states. On l.h.s of Fig.~\ref{fig2} the squared amplitude of $D^*N$-$D^*N$ transition is shown for different partial waves as function of the center-of-mass energy  for zero total momentum. In this case, the position of the $\Lambda_c(2595)$ is fitted and, as a result, the masses of the other states are slightly modified as compared to the results reported in the previous section. 

The SU(6)$\times$HQSS model in the $I=0,1$, $J=1/2,3/2$ sectors predicts several states that can have experimental confirmation:  ($I=0$, $J=1/2$) $\Lambda_c(2595)$,
($I=1$,$J=1/2$) $\Sigma_c(2823)$ and $\Sigma_c(2868)$, ($I=0$,
$J=3/2$) $\Lambda_c(2660)$,($I=0$, $J=3/2$) $\Lambda_c(2941)$, ($I=1$,
$J=3/2$) $\Sigma_c(2554)$ and ($I=1$, $J=3/2$) $\Sigma_c(2902)$
resonances. Several medium modifications are considered: no in-medium corrections, the inclusion of Pauli blocking on the nucleon
intermediate states at saturation density $\rho_0=0.17~ \rm{fm}^{-3}$, and the in-medium solution that takes into account Pauli blocking as well as the self-consistent inclusion of the $D$ and $D^*$
self-energies. These states are modified in mass and width depending on the strength of the coupling to meson-baryon channels
with $D$, $D^*$ and $N$ as well as the closeness to the $DN$ or $D^*N$ thresholds.

The knowledge of the in-medium modified dynamically generated excited baryon states is of extreme importance for the determination of the properties of open-charm mesons, such as $D$ and $\bar D$ mesons, in the nuclear medium. And the modification of the properties of open-charm mesons is of crucial value because of the implications for charmonium suppression \cite{Gonin:1996wn} and the possible formation of $D$-meson bound states in nuclei \cite{Tsushima:1998ru}.  

The properties of open charm in dense matter have been analysed in different schemes:  QMC schemes \cite{Tsushima:1998ru}, QCD sum-rule approaches \cite{Hayashigaki:2000es,Hilger:2011cq, Suzuki:2015est}, NJL models \cite{Blaschke:2011yv}, chiral effective models in matter \cite{Mishra:2003se} or pion-exchange approaches that incorporate heavy-quark symmetry constraints  \cite{Yasui:2012rw}.  Nevertheless, the full spectral features of the open-charm mesons in matter have been obtained in self-consistent unitarized coupled-channel models, where the intermediate meson-baryon channels are modified including medium corrections \cite{Tolos:2004yg,Tolos:2005ft,Lutz:2005vx,Mizutani:2006vq,Tolos:2007vh,GarciaRecio:2011xt,Tolos:2009nn,JimenezTejero:2011fc}. On the r.h.s. of  Fig.~\ref{fig2}  the $D$ and $D^*$ spectral functions are displayed for two densities at zero momentum. The $D$-meson quasiparticle peak mixes strongly with $\Sigma_c(2823)N^{-1}$ and $\Sigma_c(2868)N^{-1}$ particle-hole excitations, whereas the $\Lambda_c(2595)N^{-1}$ is visible in the low-energy tail. The $D^*$ spectral function incorporates the $J=3/2$ resonances, and the quasiparticle peak mixes with the  $\Sigma_c(2902)N^{-1}$ and $\Lambda_c(2941)N^{-1}$.  For both $D$ and $D^*$ mesons, the particle-hole modes smear out with density while the spectral functions broaden. 

\subsection{$D$-meson bound states in nuclei}

\begin{figure}[t]
\begin{center}
\includegraphics[width=.33\textwidth,angle =-90]{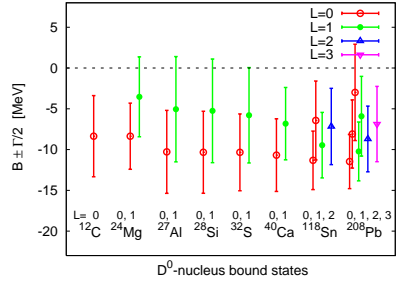}
\end{center}
\caption{$D^0$-nucleus bound states (taken from \cite{GarciaRecio:2010vt}). }
\label{figd0}
\end{figure}

\begin{figure}[t]
\begin{center}
\includegraphics[width=.48\textwidth]{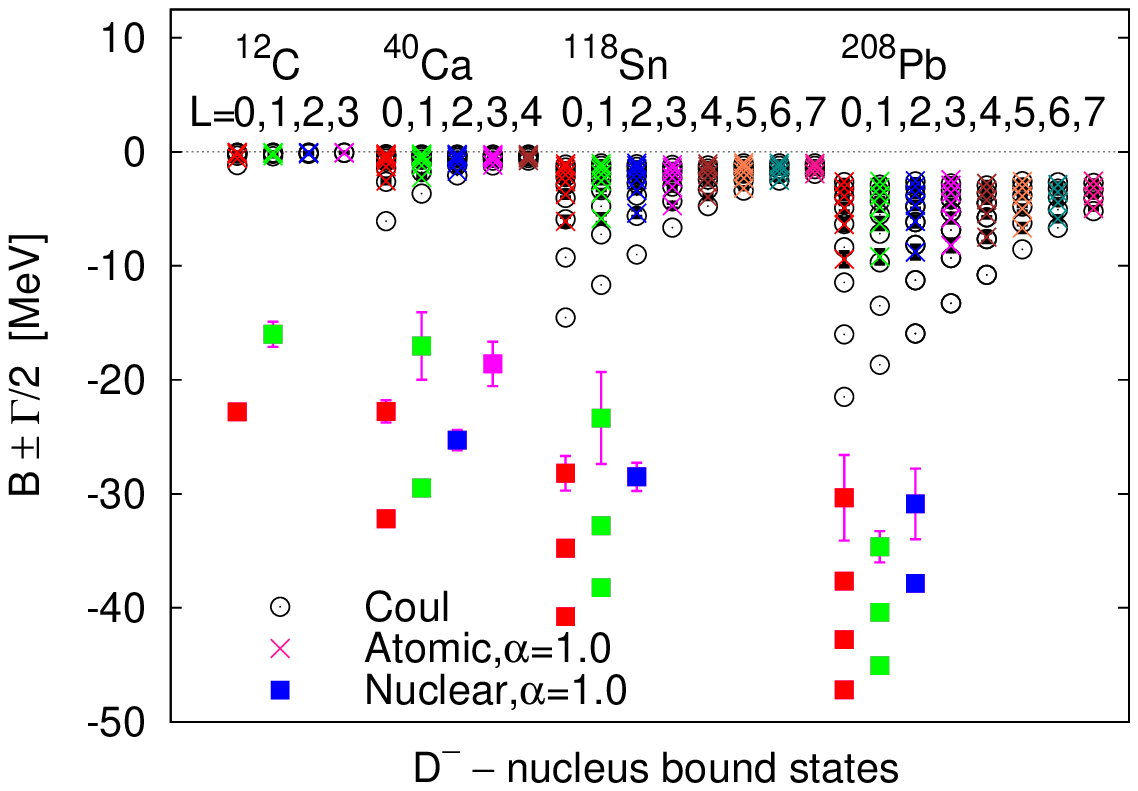}
\includegraphics[width=.48\textwidth]{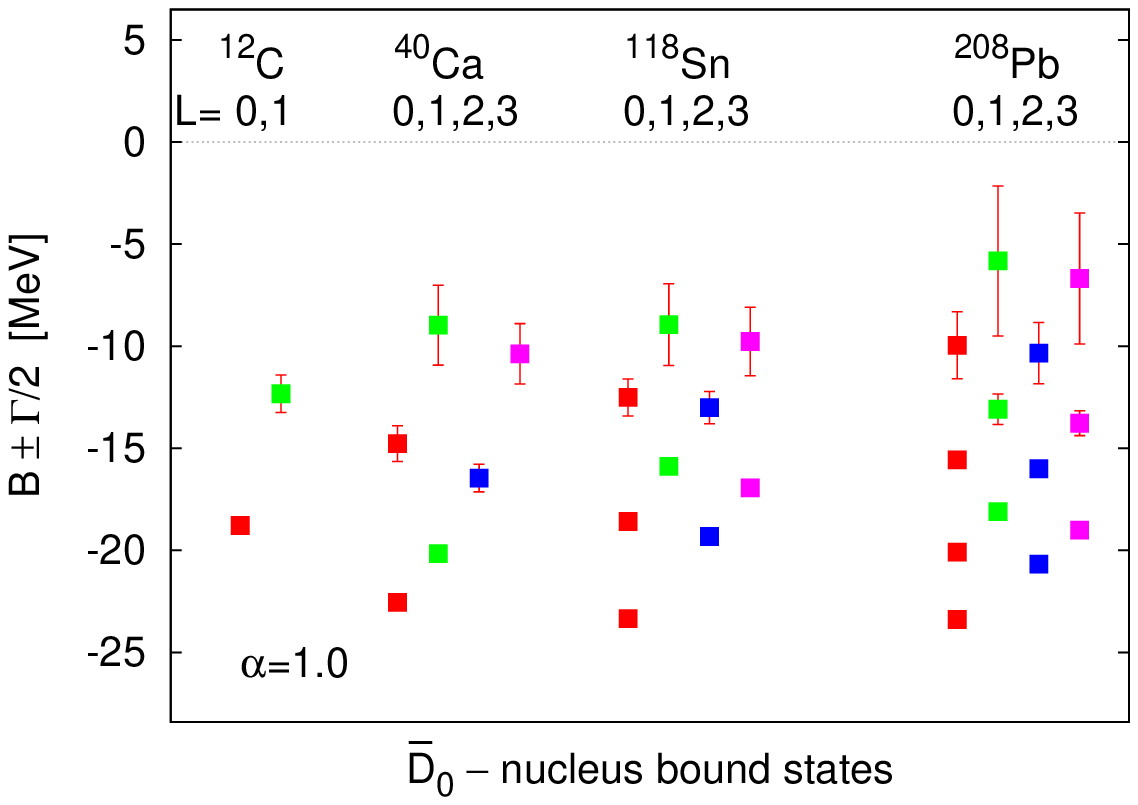}
\end{center}
\caption{$D^-$ and $\bar D^0$- nucleus bound states (taken from \cite{GarciaRecio:2011xt}).}
\label{figdm}
\end{figure}

Since the work of Ref.~\cite{Tsushima:1998ru}, there have been speculations about the formation of $D$-meson bound states in nuclei, which are based on the assumption of an attractive interaction between $D$-mesons and nucleons.  Within the model of Ref.~\cite{GarciaRecio:2010vt}, $D^0$-mesons bind in nucleus but very weakly (see Fig.~\ref{figd0}), in contradistinction to \cite{Tsushima:1998ru}. Moreover, $D^0$-mesic states show significant widths.  No $D^+$-nuclear states are found since the Coulomb interaction prevents their formation.  As for  $D^-$ or $\bar{D}^0$,  both mesons bind in nuclei as seen in Fig.~\ref{figdm}, though only nuclear states are manifest for $\bar{D}^0$ in nuclei. The atomic states for $D^-$ are less bound as compared to the pure Coulomb levels, whereas the nuclear ones are more bound and might show a significant width, appearing only for low angular momenta \cite{GarciaRecio:2011xt}.  These results are close to \cite{Yasui:2012rw}, but  in contrast to \cite{Tsushima:1998ru} for $^{208}$Pb. 

As shown in Ref.~\cite{Yamagata-Sekihara:2015ebw}, the experimental detection of $D$ and $\bar D$-meson bound states is complicated. Reactions of the type ($\bar p$, D+N) or ($\bar p$, D+2N) may be indeed possible at PANDA with antiproton beams, as long as formation cross sections are not suppressed as well as small or even zero momentum transfer reactions are feasible. More successful mechanisms could involve the emission of pions by intermediate $D^*$ or $\bar D^*$, while the resulting open-charm mesons are trapped by the nucleus \cite{Yamagata-Sekihara:2015ebw}.

\subsection{$D$-meson propagation in hot matter}
\label{propcharm}

The transport coefficients of $D$ mesons in the hot dense medium created in heavy-ion collisions offer the possibility to analyse the interaction of $D$ mesons with light mesons and baryons. Using the Fokker-Planck description,  the drag ($F_i$) and diffusion coefficients ($\Gamma_{ij}$) of $D$ mesons in hot dense matter can be obtained using an effective field theory that incorporates both the chiral and HQSS in the meson \cite{Abreu:2011ic} and baryon sectors \cite{Tolos:2013kva}.

The spatial diffusion coefficient $D_x$, that appears in Fick's diffusion law, is a relevant quantity that involves both the drag and diffusion coefficients. Within an isotropic bath, the spatial diffusion coefficient reads
\begin{equation}
\label{eq:dx} 
D_x= \lim_{p \rightarrow 0} \frac{\Gamma (p)}{m_D^2 F^2 (p)} \ , 
\label{eq}
\end{equation}
as a function of the scalar $F(p)$ and $\Gamma(p)$ coefficients. The interest on the spatial diffusion coefficient relies on the fact that it might show an extremum around the transition temperature between the hadronic and QGP phases, as seen previously for the shear and bulk viscosities \cite{Tolos:2013kva}.

\begin{figure}[t]
\begin{center}
\includegraphics[width=0.5\textwidth]{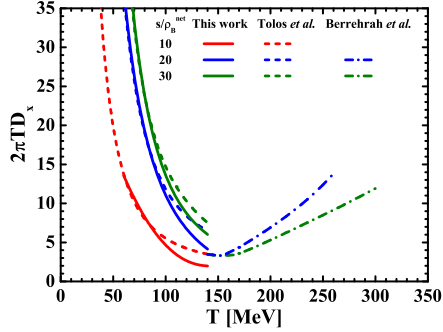}
\caption{The coefficient $2\pi T D_x$ around the transition temperature (taken from Ref.~\cite{Ozvenchuk:2014rpa}).} \label{fig5}
\end{center}
\end{figure}

In Fig.~\ref{fig5} the $2\pi T D_x$ is shown around the transition temperature following isentropic trajectories ($s/n_B$=ct) from RHIC to FAIR energies. The  matching between curves in both phases for a given value of $s/n_B$  seems to indicate the possible existence of a minimum in the $2\pi T D_x$ at the phase transition \cite{Ozvenchuk:2014rpa,Berrehrah:2014tva}.

\section{Beauty under Extreme Conditions}
\label{beauty}

\subsection{Excited beauty baryons}

The LHCb Collaboration has observed two narrow baryon
resonances with beauty, being their masses and decay modes consistent
with the quark model orbitally excited states $\Lambda_b(5912)$ and
$\Lambda^*_b(5920)$, with  $J^P=1/2^-$ and $3/2^-$,
respectively \cite{Aaij:2012da}. 

\begin{figure}[t]
\begin{center}
\includegraphics[width=0.6\textwidth]{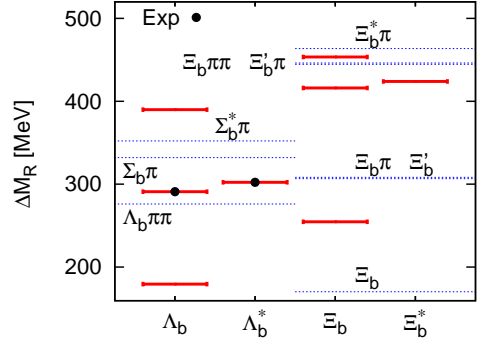}
\caption{Summary of the new predicted $\Lambda_b ^{(*)}$ and $\Xi_b^{(*)}$ states (red lines). We also show the experimentally
observed $\Lambda_b^0(5912)$ and $\Lambda_b^0(5920)$ states (black dots) and some relevant hadronic
thresholds (blue dotted lines). These results have been taken from \cite{GarciaRecio:2012db}. }
\label{lambdab}
\end{center}
\end{figure}

The existence of these states is predicted within the unitarized meson-baryon coupled-channel dynamical model, which
implements HQSS and has been presented in Sec.~\ref{charm}.  A summary of the predictions is graphically shown in
Fig.~\ref{lambdab}.   Within that scheme, the experimental $\Lambda^{(*)}_b$ states are identified as HQSS partners, 
explaining their approximate mass degeneracy.  An analogy is found between the
bottom, charm and strange sectors, given that the $\Lambda^0_b(5920)$ is
the bottomed counterpart of the $\Lambda^*(1520)$ and
$\Lambda^*_c(2625)$ states. Moreover, the $\Lambda^0_b(5912)$ belongs
to the two-pole structure similar as the one seen in the
case of the $\Lambda(1405)$ and $\Lambda_c(2595)$.

Mass and decay modes are also predicted for some $\Xi_b(1/2^-)$ and
$\Xi_b(3/2^-)$, that belong to the same SU(3)
multiplets as the $\Lambda_b(1/2^-)$ and $\Lambda_b(3/2^-)$.  Three $\Xi_b(1/2^-)$ and one $\Xi_b(3/2^-)$ states are obtained coming from the most
attractive SU(6) $\times$ HQSS representations. Two of these states,
$\Xi_b(6035.4)$ and $\Xi_b^*(6043.3)$, form a HQSS doublet
similar to that of the
experimental $\Lambda_b(5912)$ and $\Lambda^*_b(5920)$. Nevertheless, none of these states have been detected yet.

\subsection{$\bar B$ meson and $\Lambda_b$ propagation in hot matter}

The physical observables in heavy-ion collisions, such as particle ratios, $R_{AA}$ or $v_2$, are strongly correlated to the behavior of 
the transport properties of heavy hadrons. The transport properties depend on the interactions of the heavy particles with
the surrounding medium, and these are described by means of effective theories that incorporate the heavy degrees of freedom.

Following the initial works of Refs.~\cite{Abreu:2011ic,Laine:2011is,He:2011yi,Ghosh:2011bw,Das:2011vba}, the effective interaction of heavy mesons, such as $D$~\cite{Tolos:2013kva} and $\bar B$~\cite{Torres-Rincon:2014ffa} mesons, with light mesons and
baryons has been obtained by exploiting chiral and heavy-quark symmetries (see Sec.~\ref{excitedcharm} for the case of baryons with charm). With these interactions,  the heavy-meson transport coefficients  are obtained as a function of
temperature and baryochemical potential of the hadronic bath using the Fokker-Planck equation, as described in Sec.~\ref{propcharm}.

\begin{figure}[t]
\centering
\includegraphics[width=0.45\textwidth]{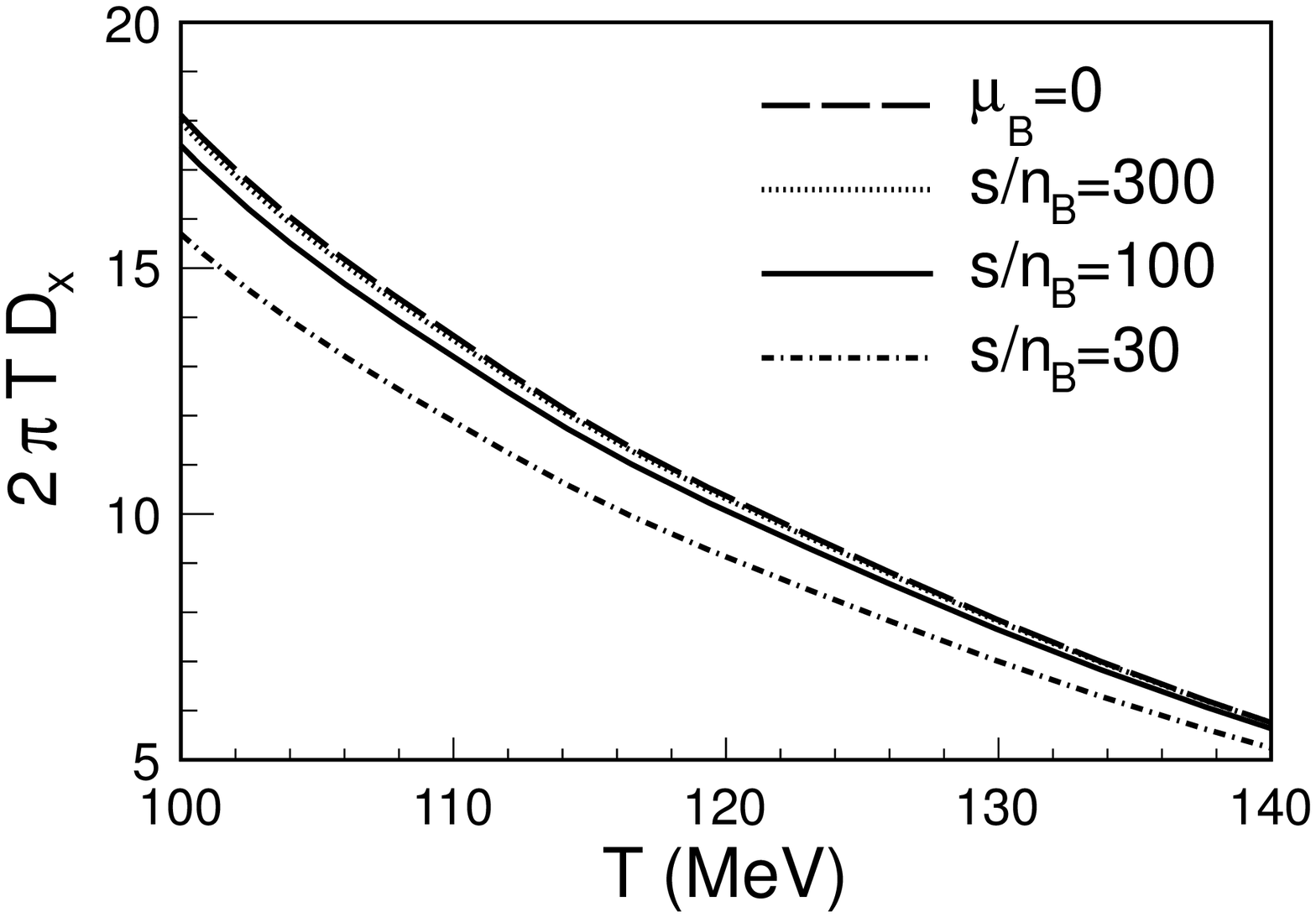}
\hfill
\includegraphics[width=0.45\textwidth]{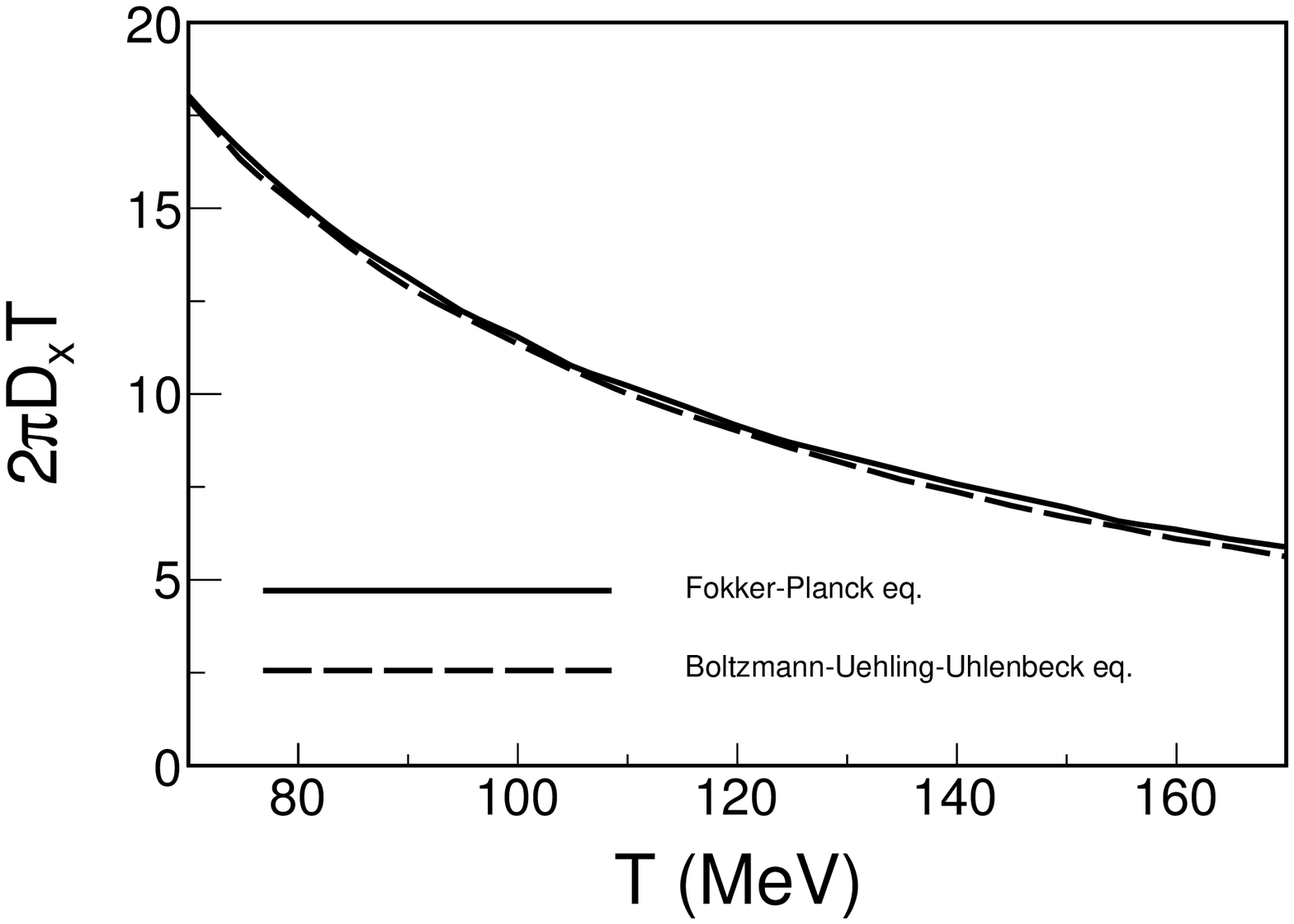}
\caption{\label{fig:diff}
Spatial diffusion coefficient, $D_x$, multiplied by the thermal wavenumber  ($2 \pi T$),  for $\bar B$ meson (left, taken from \cite{Torres-Rincon:2014ffa} ) and $\Lambda_b$ (right, taken \cite{Tolos:2016slr}). While the $\bar B$ spatial diffusion coefficient is shown for different isentropic trajectories, only the $\mu_B=0$ case is presented for $\Lambda_b$. }
\end{figure}

In Fig.~\ref{fig:diff} the spatial diffusion coefficients $D_x$, multiplied by the thermal wavenumber ($2 \pi T$), for $\bar B$ and also $\Lambda_b$ are presented, as derived in Eq.~\ref{eq}. On the l.h.s the spatial diffusion coefficient for $\bar B$ is shown for different isentropic trajectories. The results are quite independent of the entropy per baryon as long as it is high enough, that is, the collision energy
is sufficiently high. Thus, these results can be taken as prediction for the hadronic medium created at 
high energy collisions (like those at the RHIC or the LHC), independently of the precise value of the entropy per baryon of the trajectory. Although the relaxation time is smaller with larger baryonic density, the $\bar B$ meson can hardly relax to the equilibrium. Moreover, on the r.h.s of Fig.~\ref{fig:diff}  the analogous coefficient for $\Lambda_b$ is shown, but only for $\mu_B=0$. It is obtained that the outcome for the spatial diffusion
coefficient from the Fokker-Planck formalism is in good agreement to the one coming from the solution of the Boltzmann-Uehling-Uhlenbeck transport equation. Moreover, a similar behavior for both the $\bar B$ and $\Lambda_b$ spatial diffusion coefficients is observed at $\mu_B=0$, due to the comparable mass and cross sections \cite{Tolos:2016slr}. The phenomenological implications of these findings in heavy-ion collisions have been analysed in \cite{Das:2016llg}.
 

\begin{acknowledgement}
The author warmly thanks Santosh K. Das, Carmen Garcia-Recio, Juan Nieves, Lorenzo L. Salcedo, Olena Romanets and Juan M. Torres-Rincon for their collaboration and discussions that has made  this work possible. She also acknowledges support from the Heisenberg Programme of the Deutsche Forschungsgemeinschaft under the Project Nr. 383452331,  the Ram\'on y Cajal research programme, FPA2013-43425-P and FPA2016-81114-P Grants from MINECO, and THOR COST Action CA15213.
\end{acknowledgement}
\bibliographystyle{spphys}
\bibliography{mybib}
\end{document}